\documentclass[twocolumn 
]{revtex4-1}
\usepackage{amssymb}
\usepackage{graphicx}
\usepackage{amsmath,rotating}
\usepackage{color}

\def\ee{\end{eqnarray}}

\def\p{\partial}

\def\underline{\underline}

\def\=:{=\hspace{-.7em}\raisebox{1.1ex}{.}\hspace{.1em}\raisebox{-0.2ex}{.} }

\newcommand {\beq}{\begin{eqnarray}}
\newcommand {\eeq}{\end{eqnarray}}
\newcommand {\non}{\nonumber\\}

\begin{document}

\begin{flushright}YGHP-17-04\end{flushright}

\title{Confinement of Half-quantized Vortices 
in Coherently Coupled Bose--Einstein Condensates: 
Simulating Quark Confinement in QCD
}

\author{Minoru Eto$^1$, Muneto Nitta$^2$}

\affiliation{%
$^1$Department of Physics, Yamagata University, 
Kojirakawa-machi 1-4-12, Yamagata,
Yamagata 990-8560, Japan, \\
$^2$Department of Physics, and Research and Education Center for Natural
Sciences, Keio University, Hiyoshi 4-1-1, Yokohama, Kanagawa 223-8521, Japan
}%

\date{\today}

\begin{abstract}

We demonstrate that 
the confinement of half-quantized vortices (HQVs) 
in coherently coupled Bose--Einstein condensates (BECs)  
simulates certain aspects of the confinement  in 
$SU(2)$ quantum chromodynamics (QCD) in 2+1 space-time dimensions. 
By identifying the circulation of superfluid velocity as the baryon number 
and the relative phase between two components as 
a dual gluon,
we identify HQVs in a single component 
as electrically charged particles with a half baryon number. 
Further, we show that  
only singlet states of 
the relative phase of two components
can stably exist 
as bound states of vortices, that is,
a pair of vortices in each component (a baryon) 
and a pair of a vortex and an antivortex in the same component 
(a meson).
We then study the dynamics of a baryon and meson;
baryon is static at the equilibrium and rotates once it deviates from 
the equilibrium, while a meson moves with constant velocity.
For both baryon and meson we verify a linear confinement 
and determine that they are broken, thus creating other baryons or mesons in the middle 
when two constituent vortices are separated by more than 
some critical distance, 
resembling QCD.

\end{abstract}

\maketitle

\newpage

\section{Introduction}

In modern elementary particle physics, 
one of the most important and difficult problems is the confinement of quarks (and gluons) in quantum chromodynamics (QCD). 
What we daily observe in nature at low energy are not 
elementary constituents, that is, quarks and gluons
but they are strongly confined to form hadrons. 
There are two types of hadrons: 
baryons, consisting of only quarks, 
and mesons, consisting of quarks and antiquarks. 
Various baryons and mesons can exist
because there are several species (called as flavors) 
of quarks, such as up ($u$) and down ($d$).
The widely accepted explanation of  the confinement is that chromo-electric flux from a quark is squeezed to form a flux tube in
a dual superconductor, 
in which magnetic monopoles are condensed 
\cite{Nambu:1974zg,tHooft,Mandelstam:1974pi}. 
Thus, the interaction energy 
between (anti-)quarks is proportional 
to the distance between them; this
is a salient signal of the confinement.  
Although the explanation  
is quite plausible, 
it is  
difficult to prove it.
Therefore, many studies have been conducted for QCD-like theories.

From among these studies, Polyakov \cite{Polyakov:1976fu} made an
important remark using a duality 
about a $U(1)$ gauge theory in 2+1 space-time dimensions;
this can be obtained as the low-energy limit of  
an $SU(2)$ gauge theory with a triplet scalar field.
The duality mentioned here is the one 
between the $XY$ and Abelian--Higgs models in $2+1$ dimensions \cite{Peskin:1977kp,Dasgupta:1981zz},
providing insights for understanding the fractional quantum Hall effect \cite{Lee:1989fw},
Mott transitions 
\cite{Beekman:2016}, etc.
As a photon has only one polarization in three spece-time dimensions, it can be dualized to 
the so-called dual photon
(a periodic scalar field) $\theta \in [0,2\pi)$ 
defined through
$
 \p_\mu A_\nu - \p_\nu A_\mu 
= \frac{g_c^2}{4\pi} \varepsilon_{\mu\nu\rho}\p^\rho\theta$
with a coupling $g_c$.
Under the duality relation, electrically charged bosons in the original theory are interchanged by vortices in $\theta$.
The dual photon is  massless in the perturbation theory but it gets mass 
$m_{\rm dp}$ 
through nonperturbative monopole effects,
and is consequently described at low energy by the Lagrangian
\beq
{\cal L}_{\rm dp} = \frac{g_c^2}{32\pi^2}\p_\mu\theta\p^\mu\theta + c g_c^2 m_{\rm dp}^2 \cos \theta
\label{eq:Ldual}
\eeq
where $c$ is a constant. 
Clearly, a $\theta$ vortex 
must be attached by 
a soliton because of the potential term, 
which corresponds to an electric flux tube in the original theory
showing confinement.

In this work, 
we show that this confinement phenomenon 
can be simulated in ultracold atomic gases 
of coherently coupled 
two-component BECs, which is 
an experimentally
controllable ideal system
\cite{Dalfovo:1999zz,Pitaevskii:2003,Pethick:2008},
as realized by the JILA group \cite{Hall:1998,coherent}. 
In particular,
quantized vortices in ultracold atomic BECs have been
studied thoroughly since their experimental realization \cite{Fetter}.
The recent progress in this subject can be seen in the development
of techniques to nucleate the vortices and
detect real-time vortex dynamics \cite{vortex-exp}.
Theoretically, 
they are described by
the Gross--Pitaevskii (GP)
equations:
\beq
\!\!\!\!\!\!\!\!
\left[  i\hbar \frac{\p}{\p t}
\!+\! \frac{\hbar^2}{2m}\nabla^2 \!- \! \left(g_i |\Psi_i|^2 \!+\! g_{12}|\Psi_{\hat i}|^2 \!-\!\mu_i \right)\!\right]\!\! \Psi_i
\!=\! -  \hbar \omega \Psi_{\hat i}\,,
\label{eq:gp1}
\eeq
where $\hat 1 = 2$, $\hat 2 = 1$, 
$g_{ij}$ represents the atom--atom coupling constants, $m$ is the mass of atom, and $\mu_i$ represents the chemical potential. 
The first and second condensates $\Psi_{1,2}$ are coherently coupled
through the Rabi (Josephson) terms with the Rabi frequency $\omega$. 
Such coherent coupling was achieved by the JILA group \cite{coherent}.
In the following, we assume $g_1 = g_2 = g$ and $\mu_1 = \mu_2$ for simplicity, and focus on
a miscible BEC ($g > g_{12}$)
in which both condensates coexist as
$
v = |\Psi_1| = |\Psi_2| = \sqrt{(\mu + \hbar \omega)/(g + g_{12})}.
$
The system has the symmetry $[U(1)_S \times U(1)_R]/\mathbb{Z}_2$:
$
(\Psi_1,\Psi_2) \to (e^{i\alpha}\Psi_1,e^{\pm i\alpha}\Psi_2)\,,
$
where $+$ is for $U(1)_S$ and $-$ is for $U(1)_R$. 
The Lagrangian for Eq.~(\ref{eq:gp1}) is
\beq
{\cal L}_{\rm GP} &=& \sum_i \bigg[- \frac{i\hbar}{2}\left(\Psi_i\dot \Psi_i^* - \dot\Psi_i \Psi_i^*\right) - \frac{\hbar^2}{2m}|\nabla\Psi_i|^2
 \non
&& + \mu_i |\Psi_i|^2 - \frac{g_i}{2}|\Psi_i|^4\bigg] - g_{12} |\Psi_1\Psi_2|^2 - V_R,
\label{eq:LGP}
\eeq
where $V_{\rm R} = -  \hbar \omega \left(\Psi_1\Psi_2^* + \Psi_1^*\Psi_2\right)$. 
We then truncate this by
substituting the expression of the condensates
$\Psi_i = (v + r_i)e^{i\theta_i}$ into Eq.~(\ref{eq:LGP}) 
and  by integrating out the amplitudes $r_i$:
\beq
\tilde {\cal L}_{\rm GP} &=& 
\frac{\hbar^2}{4(g+g_{12})}\dot\theta_+^2 + 
\frac{\hbar^2}{4(g-g_{12})}\dot\theta_-^2- \frac{\hbar^2 v^2}{4m} (\nabla \theta_+)^2\non
&&
- \frac{\hbar^2 v^2}{4m} (\nabla \theta_-)^2
+ 2\hbar \omega v^2 \cos \theta_-,
\label{eq:rLGP}
\eeq
where we ignored constants and managed the Rabi term perturbatively. 
Here, $\theta_+ \equiv \theta_1+\theta_2$ is a phonon 
and $\theta_- \equiv \theta_1-\theta_2$ is known 
as the Leggett mode or phason, corresponding 
to $\theta$ in Eq.~(\ref{eq:Ldual}). 
The superfluid velocity is $\mathbf{v}_{\rm s} = \frac{\hbar}{m}\nabla \frac{\theta_1 + \theta_2}{2}$, and vortex winding number is defined by 
$
N_B = 
\frac{m}{h}
\oint_C d\mathbf{s} \cdot \mathbf{v}_{\rm s} = \frac{n_1 + n_2}{2}$,
where  $C$ is a closed path enclosing all the vortices, and $n_i \in \mathbb{Z}$. 
In the miscible ground state,
vortices winding in $\theta_1$ and $\theta_2$ 
are half-quantized vortices (HQVs);
this significantly differs from the simpler model 
in Eq.~(\ref{eq:Ldual}). 
HQVs have been observed recently \cite{Seo}.
Our proposal is to identify the winding number $N_B$  
as a baryon number in $SU(2)$ QCD, and 
HQVs winding in $\theta_1$ and $\theta_2$ 
as $u$ (up) and $d$ (down) (bosonic) quarks, respectively. 
Previous studies suggested \cite{Son:2001td} 
and confirmed \cite{Kasamatsu:2004tvg,Kasarev}
that 
$u$- and $d$-vortices are connected through 
a soliton resembling the confinement 
(see also \cite{Garcia:2002,Cipriani:2013nya,Calderaro:2017}).
In Ref.~\cite{Tylutki:2016mgy}, 
a similarity based on 
disintegration of confining strings  
between HQVs 
was pointed out. 
In the current study, we show that 
all stable bound states of vortices are singlet states of $U(1)_R$,
that is, 
a baryon $ud$ and mesons  $\bar u u$ and $\bar dd$.
We further identify $U(1)_R$ as 
a dual gluon (the diagonal component of $SU(2)$ color symmetry).
We then achieve a fine agreement with $SU(2)$ QCD 
for which confinement allows only singlet states of $SU(2)$.
We further numerically simulate 
the disintegration of solitons 
in baryon and meson when 
constituents are separated, as in the QCD case. 


\section{Confinement}

\paragraph{Liberated vortices} 
The static intervortex forces between the well-separated unconfined HQVs ($\omega = 0$) at distance $R$ were found \cite{Eto:2011wp} as
$\frac{U_{uu}}{U_*} =  -   \frac{2}{\pi} \log \rho$ and  
$\frac{U_{ud}}{U_*}  = \frac{\pi}{2} \frac{g_{12}}{g-g_{12}} \frac{1}{\rho^2} \log \frac{\rho}{\xi}$,
with $U_* = v^2 \hbar^2/m$, $R_* = \sqrt{2\hbar^2/\mu m}$, $\rho = R/R_*$, and $\xi$ is a dimensionless constant. 
The interactions between other pairs are $U_{uu} = U_{dd} = - U_{\bar u u} = - U_{\bar d d}$ and
$U_{ud} = U_{\bar u d} = U_{\bar d u} = U_{\bar u \bar d}$.
Therefore, irrespective of the sign of the coupling constants $g$ and $g_{12}$, 
$U_{uu,dd}$ is repulsive while $U_{u\bar u,d\bar d}$ is attractive. They are nothing but
the 2+1 dimensional Coulomb potentials.
In contrast, the potentials between the different species are blind to whether 
they are vortex or antivortex, and are always repulsive (attractive) for $g_{12} > 0$ ($g_{12} < 0$).

By applying the point particle approximation to Eq.~(\ref{eq:gp1}) \cite{Kasamatsu:2015cia}, we can derive 
$
\frac{d Z_i}{d\tau} = - \frac{i}{\pi q_i} \frac{\p}{\p \bar Z_i} \frac{U}{U_*}, 
$
where 
$Z_i(\tau) = X_i(\tau) + i Y_i(\tau)$ is the 
position of the $i$-th vortex in terms of the dimensionless coordinates $\tau = \mu t/2\hbar$ and $\vec y = \vec x/R_*$, and
$U$ is either of the $U_{\bar uu}$ or $U_{ud}$ type according to components of the pair.
We use the dimensionless coupling $u = 2g/\mu$,
$u_{12} = 2g_{12}/\mu$, and
$\eta = 2\hbar\omega/\mu$ in the following.
As $U$ depends only on $R/R_*= |Z_1 - Z_2|$, this can be easily solved:
Each component of the $uu$ ($dd$) pair rotates the other counterclockwise, while
the pair $ud$ rotates counterclockwise (clockwise) for $u_{12} > 0$ ($u_{12}<0$).
As $U_{uu} \sim \log R$ is much greater than $U_{ud} \sim (\log R)/R^2$,  $uu$ ($dd$)
rotates quicker than $ud$.
The other pairs, $\bar u u$, $\bar d d$, $\bar u d$, and $\bar d u$ show linear and parallel motions, with the distance $R$ being preserved.

\paragraph{Confined vortices} 
When $\omega \neq 0$, $U(1)_R$  is explicitly broken,
metastable sine-Gordon soliton appears which is also called as a magnetic domain wall \cite{Usui:2015,Qu:2016,Qu:2016b,Calderaro:2017}. 
The solution and its tension are given as
$
\theta_-
= 4 \arctan\exp\sqrt{\frac{2m \omega}{\hbar}}\,x$,
and 
$
T_{\rm SG} = 8 v^2 \frac{\hbar^2}{m}\sqrt{\frac{m\omega}{\hbar}},
$
respectively.
This confines the constituent vortices, so that 
only $U(1)_R$ singlet composite states
(a baryon $ud$ with $N_B=1$
and a meson $\bar u u$ (or $\bar d d$) 
with $N_B=0$) must remain stable as in QCD. 
The remaining pairs, $uu$, $dd$, $\bar u d$, and $\bar d u$, 
which are not $U(1)_R$ singlets, should
 not appear \cite{footnote2}.

\begin{figure}[t]
\begin{center}
\includegraphics[width=8cm]{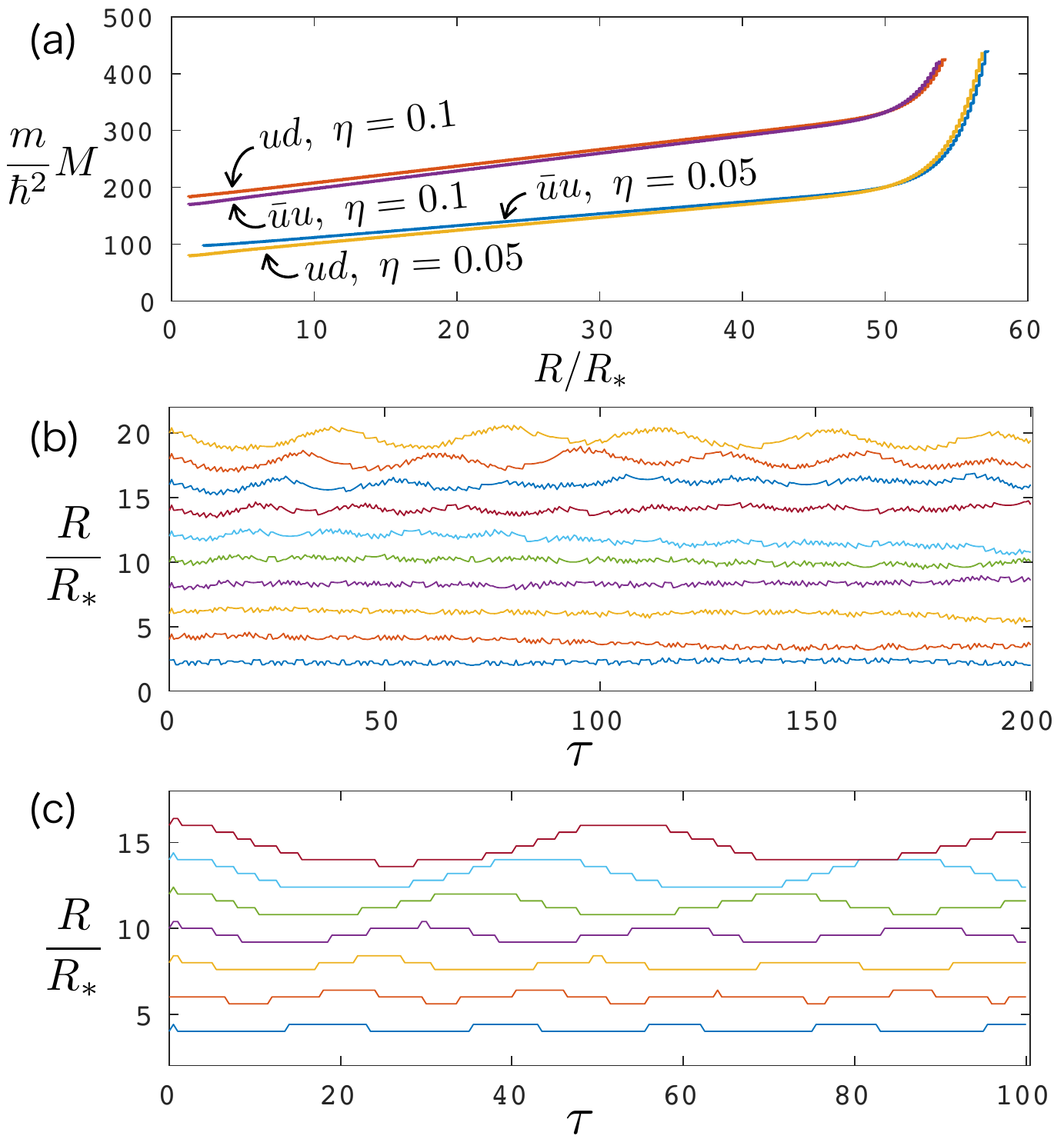}
\caption{ 
(a) Mass $M$ of a molecule of size $R$ under the relaxation process with initial separation is $60 R_*$.
The four cases are shown for the meson $\bar uu$ and baryon $ud$ with $\eta = 0.05$ and $0.1$.
Fluctuations of the baryon (b) and the meson (c) are shown under the real time evolution.  
The couplings are chosen as $u = 2 u_{12} = 1$.
}
\label{fig:linear_relax}
\end{center}
\end{figure}

In addition to the intervortex potentials $U_{ud,\bar u u}$,
the linear confining potential $U_{\rm SG} = T_{\rm SG} R$ 
contributes to the vortex dynamics.
However, note that $U_{\rm SG} = T_{\rm SG} R$ is an approximation, which 
is obtained under the assumption that $|\Psi_i| = v$ holds everywhere. 
Therefore, we should correct the soliton tension 
appearing inside the baryons and mesons.
Hence, we applied the imaginary time $\tilde t$ evolution method to an initial configuration 
made by simply superposing two unconfined vortex solutions ($\omega = 0$)
with a large initial distance $R = 60 R_*$. 
Fig.~\ref{fig:linear_relax}(a) plots the static total mass $M$ from Eq.~(\ref{eq:LGP})
as a function of molecule size $R(\tilde t)$.
Initially, $M$ rapidly decreases 
because of a large deformation in the condensate profiles, while the molecule size remains almost unchanged. After a while,
the mass reduction moderates and the imaginary evolution slows. At that instance, the function $M(R)$ exhibits linear behaviors,
as shown in Fig.~\ref{fig:linear_relax}(a). 
The tension of a soliton is presented by the slope of the linear behavior. 
We numerically confirmed that the tension is smaller than $T_{\rm SG}$ 
by approximately $10\%$--$20\%$, and the correct confining potential is
\beq
U_{\rm conf}  = \alpha T_{\rm SG} R = 8 \alpha U_* \sqrt{\eta}\, \rho.
\label{eq:true_SG_pot}
\eeq
For example, according to Fig.~\ref{fig:linear_relax}(a), 
$\alpha = 0.85$ for $ud$ while $\alpha = 0.88$ for $\bar u u$, with $\eta = 0.1$.  

Let us next turn to real time dynamics. 
Our initial configurations 
are obtained as a result of the imaginary time evolution with sufficiently long duration 
until at least the configuration reaches the linear line, as shown in Fig.~\ref{fig:linear_relax}(a). 
With this configuration as the initial configuration, we switch on the real time evolution.
The baryon $ud$ for $u_{12} > 0$ is static at an equilibrium where
the repulsive intervortex force  
is balanced with the soliton's attractive force.
Once the distance $R$ is deviated from the equilibrium distance $R_0$,
the interaction is repulsive (attractive) for $R < R_0$ ($R > R_0$),
thus causing the baryon to rotate. 
By numerically solving GP equations (\ref{eq:gp1}) we confirmed
that the baryon of $R < R_0$ rotates counterclockwise 
while that of $R > R_0$ rotates clockwise.
See Appendix A and the movies in Ancillary files 
for details of the numerical solutions to 
the GP equations.  
This can also be explained by the point particle approximation explained earlier that included the confining potential $U_{\rm conf}$.
The total effective potential $U_{\rm tot} = U_{ud} + U_{\rm conf}$ 
remains a function on $R$ only so that it can be solved easily.
The $ud$ molecule rotates (counter)clockwise when $U_{\rm tot}$ is attractive (repulsive) with
constant period ${\cal T} = 2\pi^2 \rho \left(\frac{\p}{\p \rho} \frac{U_{\rm tot}}{U_*}\right)^{-1}$. 
This works similarly for the mesons by replacing $U_{ud}$ by $U_{\bar u u}$.
They linearly move with a constant velocity ${\cal V} = \frac{1}{2\pi}\frac{\p}{\p \rho} \frac{U_{\rm tot}}{U_*}$.
The effect of the soliton can be most clearly seen when the separation is large where
${\cal T}$ and ${\cal V}$ asymptotically behave as
\beq
{\cal T}  \to  \frac{\pi^2}{4\alpha \sqrt{\eta}} \rho,\qquad
{\cal V} \to \frac{8\alpha\sqrt{\eta}}{\pi}, \qquad (R \gg R_*)\,.
\label{eq:asymptotic}
\eeq
We confirmed these asymptotic behaviors by numerically solving the GP equation.
$\alpha$ can be determined from Eq.~(\ref{eq:asymptotic}), for example $\alpha = 0.85$ for $\eta=0.1$ baryon, which is 
consistent with the result from the imaginary time evolution in Eq.~(\ref{eq:true_SG_pot}).

\section{Fragmentation of soliton}

\setcounter{paragraph}{0}

\begin{figure}[h]
\begin{center}
\includegraphics[width=7cm]{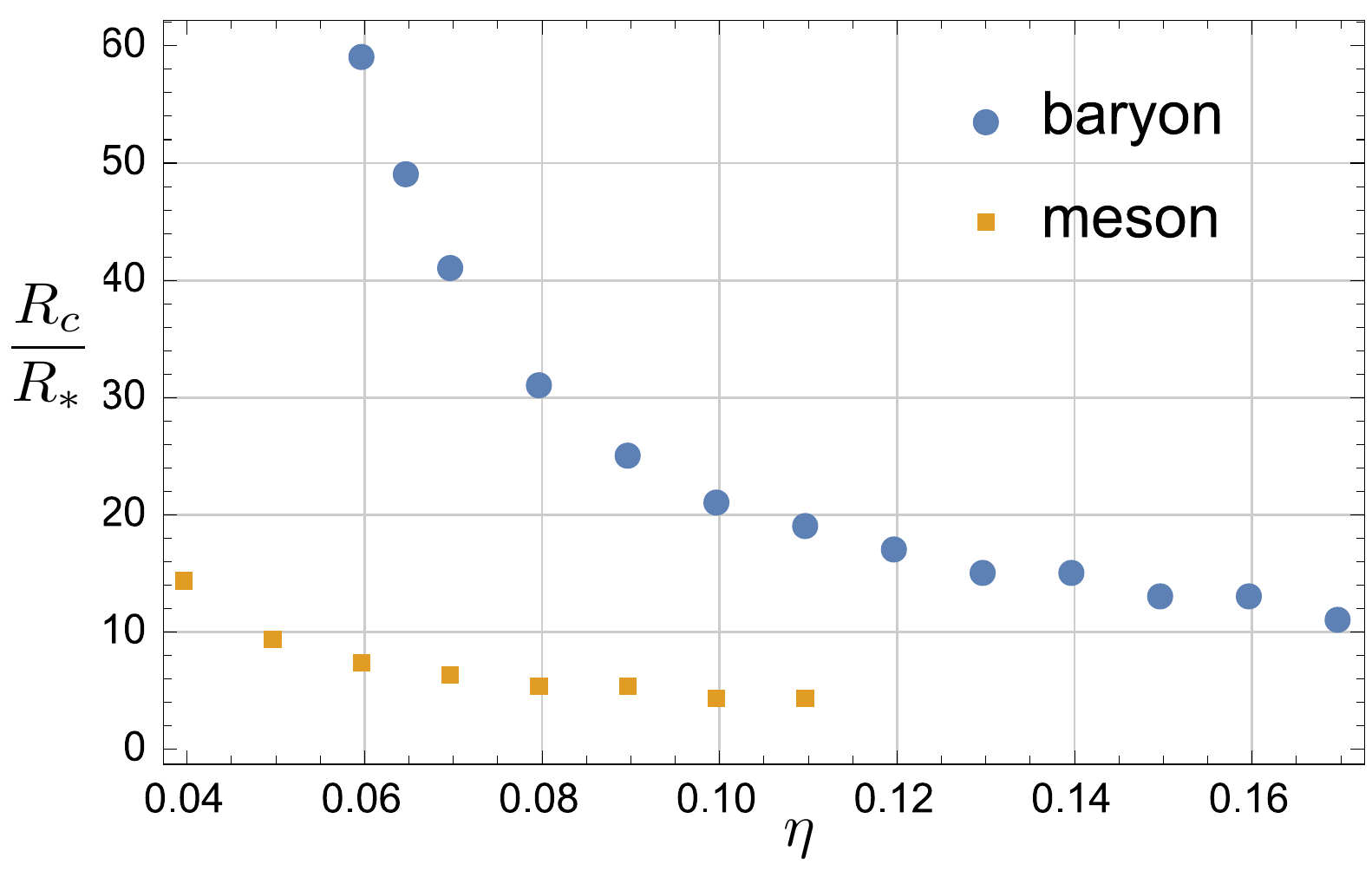}
\caption{
Relations between $\eta$ and the critical molecule size $R_{\rm c}$ at which fragmentation occurs for $u=2u_{12}=1$.
}
\label{fig:R_c}
\end{center}
\end{figure}

When a molecule is elongated beyond a critical length $R_c$, 
the soliton will break up into
small pieces. 
While approaching to the critical size $R_{\rm c}$ from a depressed size, the molecule starts to oscillate,
and the instability develops toward fragmentation.
We plot the molecule size as a function of time 
in Fig.~\ref{fig:linear_relax}(b) and (c),
which show that the larger molecule has larger amplitude in the oscillational mode. 
Fig.~\ref{fig:R_c} shows the molecule fragments when its size reaches $R_{\rm c}$.
As $T_{\rm SG}$ is proportional to $\sqrt{\eta}$, $R_{\rm c}$ recueces (enlarges) for a large (small) $\eta$.
As both $U_{\bar uu}$ and $U_{\rm conf}$ are attractive for the meson, 
$R_{\rm c}$ for the meson is smaller than that for the baryon for $u_{12} >0$.

The disintegration of 
a baryonic molecule  
in a harmonic trapping potential was reported in Ref.~\cite{Tylutki:2016mgy}.
In the current study, we investigated both baryons and mesons.
In contrast to in findings in \cite{Tylutki:2016mgy} ($u_{12} = 0$),
we consider $u_{12} >0$ and a homogeneous system without
a trapping potential 
to exclude external effects from environment and focus on 
consequences purely due to confinement.

\begin{figure}[h]
\begin{center}
\includegraphics[width=8.7cm]{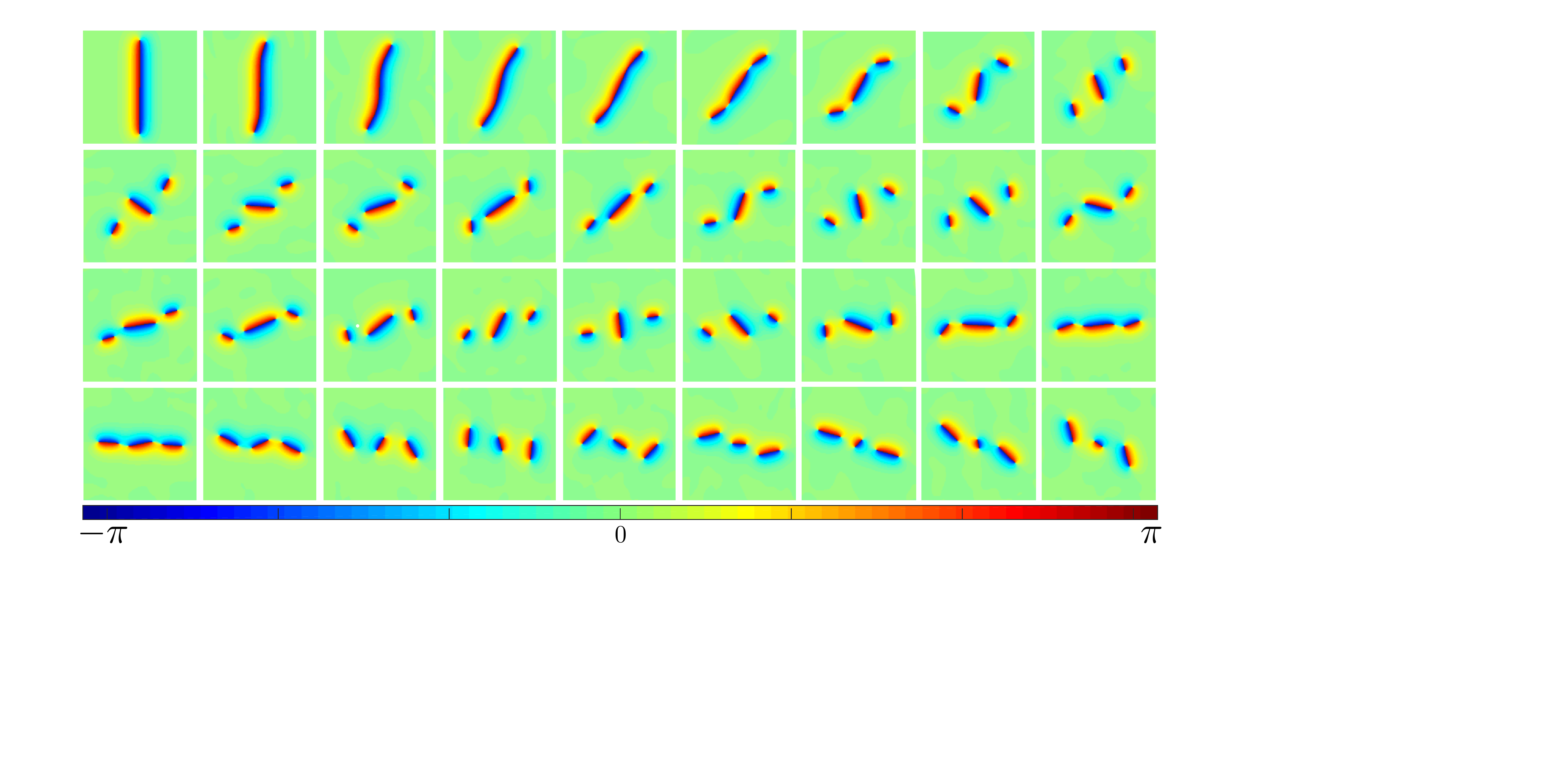}
\caption{
Motion of a $ud$ baryon ($\tau=0$\,-\,$175$ with a step $\Delta \tau = 5$) with $R_{\rm ini}/R_* = 24$. 
$\theta_-$ is plotted.
The spatial region  $\vec y \in [-15,15]^2$ is shown. $u = 2 u_{12} = 10 \eta = 1$.
}
\label{fig:baryons}
\end{center}
\end{figure}

\paragraph{Baryonic molecule}\ \  
We conducted numerous simulations for various initial molecule sizes $R_{\rm ini}$. 
To provide a bench-mark, 
we set 
$\eta = 0.1$ with    
the equilibrium length $R_0 = 0.66 R_*$
and critical length $R_{\rm c} \simeq 21 R_*$.
We set $R_{\rm ini}$ much larger than $R_0$. 
$ud$ rotates as if it is a solid stick when its size is not extremely large, for example,
$R_{\rm ini} = 8 R_*$.
When we further enlarge the molecule $(R_{\rm ini} = 20 R_*)$,  
though $ud$ does not break up, 
the soliton starts twisting sideways. 
When $R_{\rm ini}$ becomes larger than the critical value $R_{\rm c}$, 
fragmentation occurs for $R_{\rm ini} = 24R_*$, as shown in Fig.~\ref{fig:baryons}.

In detail,
the fragmentation occurs as follows. The $u$ and $d$ vortices at the edges rotate rapidly with the period given in the
asymptotic formula (\ref{eq:asymptotic}); however, the long soliton maintains its initial position. As a result, a large gap emerges between 
the rotation speed of soliton near the edge and center. 
This causes the soliton to curve and break up. 
The fragmentation always occurs at points near the edges. 
After short baryons are released from the edge, an antibaryon remains at the center. As shown in Fig.~\ref{fig:baryons},
the two baryons at the outer sides rotate clockwise, and simultaneously revolve clockwise. 
The revolution speed of the two outer baryons is much slower than the rotation speed of a long original baryon before breaking up.
This is because after the fragmentation, the two short baryons at outer sides are no longer connected by the soliton,  reducing
the interaction, and therefore the rotation speed is very small.
However, the antibaryon at the center rotates counterclockwise.
Interestingly, the lengths of baryon and antibaryon oscillate from time to time. 
If we place an even longer baryon initially, it splits up into smaller pieces, 
see Appendix A.

\begin{figure}[t]
\begin{center}
\includegraphics[width=8.7cm]{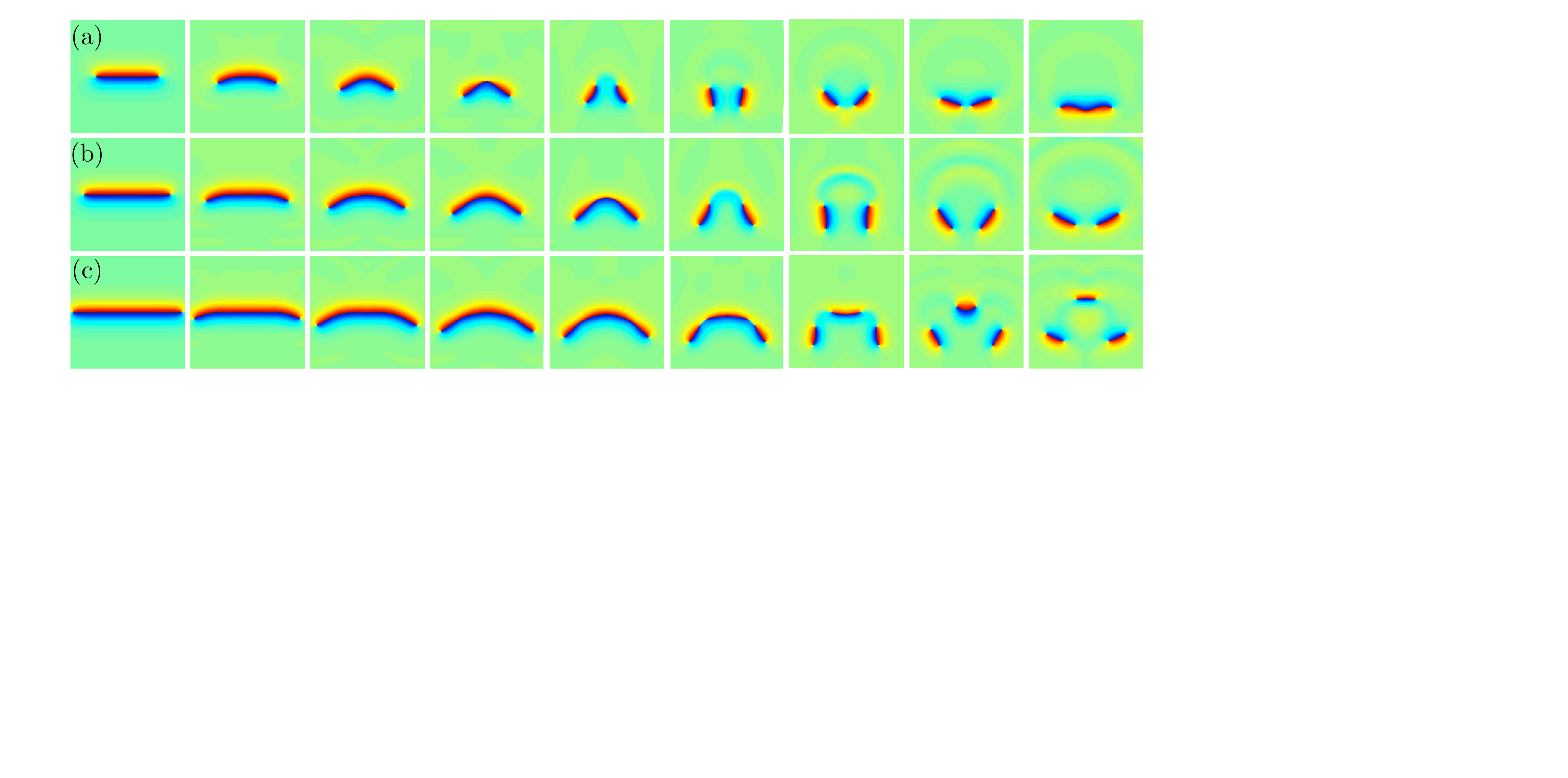}
\caption{
(a)-(c) 
Motions of a meson
($\tau = 0$\,-\,$40$ with a step $\Delta \tau =5$)
with $R_{\rm ini}/R_* = 16,22,28$, respectively. 
See the caption of Fig.~\ref{fig:baryons} for the parameters.
}
\label{fig:long_meson}
\end{center}
\end{figure}

\paragraph{Mesonic molecule}\ \ 
For concreteness, we again set $\eta = 0.1$. 
We initially prepared $\bar u u$.
A short meson runs downward with a constant speed as a solid stick.
A slightly longer meson, shown in the panels of the line Fig.~\ref{fig:long_meson}(a) for $R_{\rm ini}/R_* = 16$
breaks up into a pair of baryon and antibaryon by creating $\bar d$ and $d$ at the middle, as $\bar u u \to \bar u \bar d +d u$. 
The pair continues to run straight 
downward as $\bar u \bar d$ and $d u$ can be seen as a pair of integer vortex and antivortex, respectively, 
while they rotate oppositely. After a while, when 
they rotate by $180^\circ$, they fuse again and become another type of meson, namely $\bar d d$.
The meson repeats this exchange $\bar u u \leftrightarrow \bar dd$.
Interestingly, this suggests that the actual meson state 
is $\bar u u \pm \bar dd$ (in a long period), as in QCD.
The longer $\bar u u$ meson ($R/R_*=22$), as shown in panels of the line Fig.~\ref{fig:long_meson}(b), 
also fragments and changes as $\bar u u \to \bar u \bar d +d u$.
Unlike case (a), $du$ and $\bar u \bar d$ do not coalesce, and an up-going rarefaction pulse is emitted. 
The pair of baryon and antibaryon rotate oppositely and run downward.
A very long meson with $R/R_*=28$, shown in panels of the line Fig.~\ref{fig:long_meson}(c), is quite similar to that in case (b),
except that it now emits another meson instead of a rarefaction pulse. The initial meson breaks up into three pieces 
as $\bar u u \to \bar u \bar d+d \bar d+d u$. The created meson at the middle is $d \bar d$ not $\bar d d$; therefore, it runs upward.
If we prepare a much longer meson initially, it will break up into larger number of small molecules, see Appendix A and movies in 
Ancillary files.

\section{Concluding remarks}

In this work, we  studied the real time dynamics 
and  confinement of 
baryonic and mesonic vortex molecules in 
coherently coupled 
two-component BECs  
in 2+1 dimensions. 
We observed that 
only $U(1)_R$ singlet states of vortex bound states can appear
linearly confined by solitons, thereby simulating QCD.  
In addition, we numerically showed
that a short baryon rotates and a short meson moves straight; this can be accounted well by the point vortex approximation, 
whereas molecules longer than $R_c$ involve fragmentation.
The disintegration of mesons includes rich phenomena, that is, a meson transitioning into a pair of a baryon and an antibaryon, 
flavor oscillation,
emission of rarefaction pulses, and/or oppositely oriented mesons.

A lattice of baryonic molecules was 
constructed in the rotating BEC \cite{Cipriani:2013nya}.
Real time dynamics of a few (and many) body system 
will be an important direction.
As real QCD has  $SU(3)$ symmetry, our next step will be  
the confinement of 1/3 quantized vortices in 
three-component BECs,
for which a baryon was constructed 
numerically in Ref.~\cite{Eto:2012rc,Nitta:2013eaa}.
While we have shown confinement of charged bosons, 
the recently highlighted duality between vortices and 
fermions \cite{Son:2015xqa}
may provide simulation of QCD with quarks (as fermions). 

\ \\

{\bf Acknowledgments}\ \ 
This work is supported by the Ministry of Education,
Culture, Sports, Science (MEXT)-Supported Program for the Strategic
Research Foundation at Private Universities ``Topological Science''
(Grant No.~S1511006).
This work is also supported in part by the Japan Society for the Promotion of Science (JSPS) 
Grant-in-Aid for Scientific Research (KAKENHI) Grant Numbers (26800119 (M. E.) and 16H03984 (M. E. and M. N.)).
The work of M.~N.~is also supported in part by a Grant-in-Aid for
Scientific Research on Innovative Areas ``Topological Materials
Science'' (KAKENHI Grant No.~15H05855) 
and ``Nuclear Matter in Neutron
Stars Investigated by Experiments and Astronomical Observations''
(KAKENHI Grant No.~15H00841)
from the MEXT of Japan.

\appendix

\section{Numerical details}

\begin{figure}[h]
\begin{center}
\includegraphics[width=8cm]{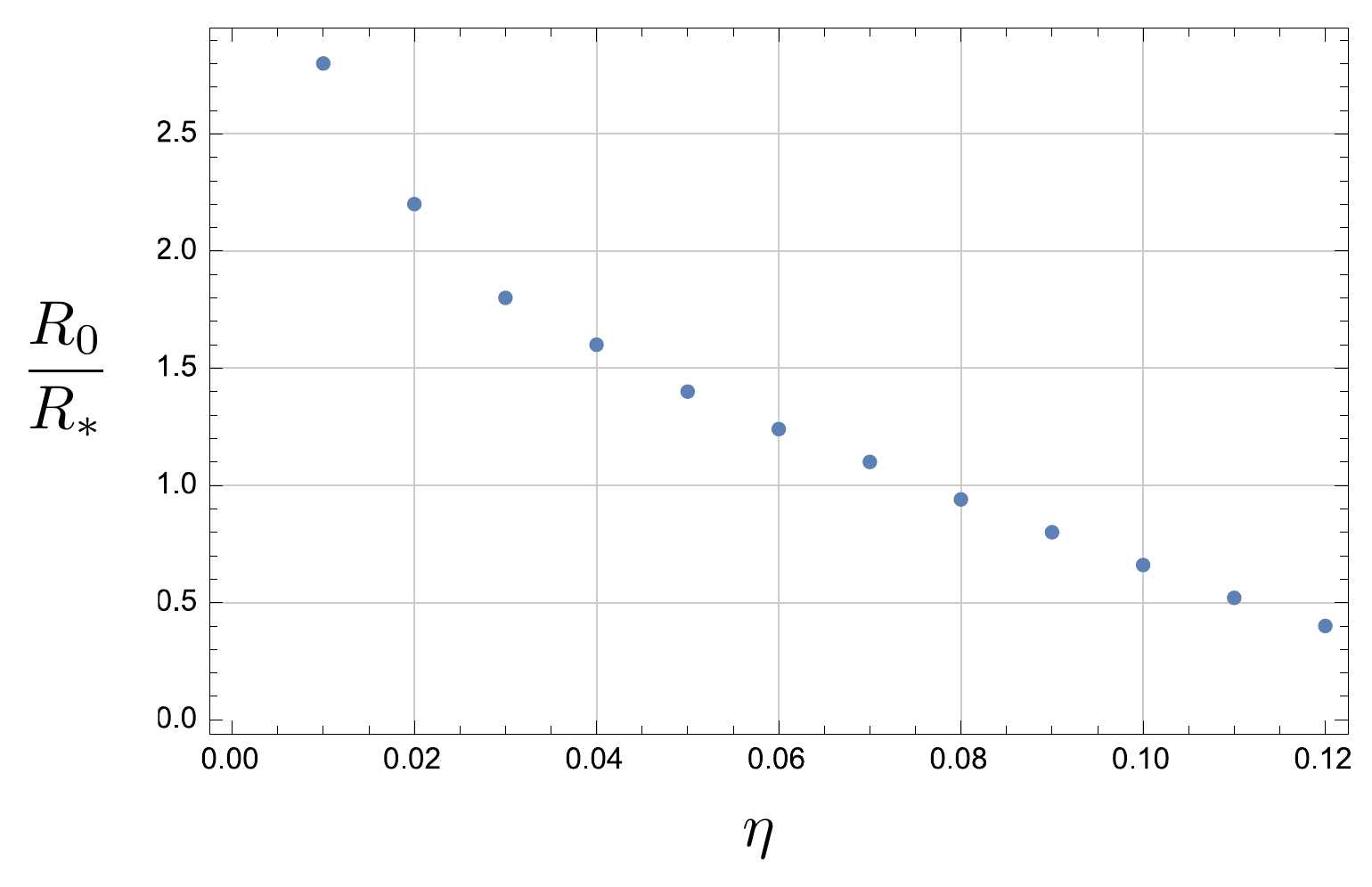}
\caption{
The equilibrium size $R_0$ of the baryonic molecule $(ud)$ as function of $\eta$.
The couplings are chosen as $u = 2 u_{12} = 1$.
}
\label{fig:baryon_size}
\end{center}
\end{figure}

The baryonic molecule for $g_{12} > 0$ at the equilibrium is static because 
the inter-vortex repulsion is canceled by the confinement force by the soliton.
The equilibrium distance $R_0$ depends on
the Rabi coupling $\eta$ as shown in Fig.~\ref{fig:baryon_size}.
As expected, the baryon shrinks as we increase the Rabi frequency 
$\eta$ since
the soliton tension is proportional to 
$\sqrt{\eta}$.
\begin{figure}[h]
\begin{center}
\includegraphics[width=8cm]{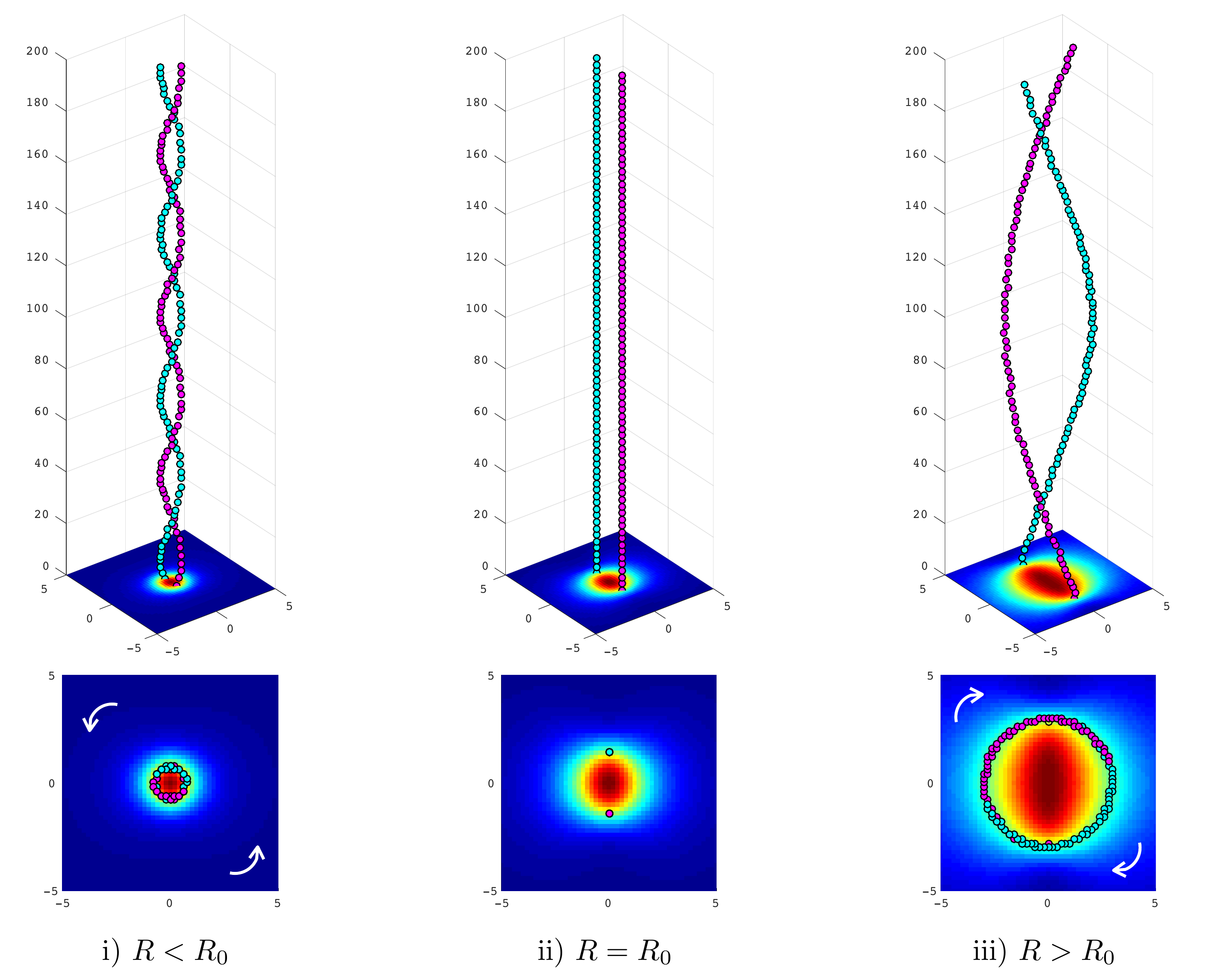}
\caption{
Typical motions of the $ud$ baryon with the separation i) $R<R_0$, ii) $R = R_0$, and iii) $R > R_0$
with $R_0 = 2.8R_*$.
The density plots at the bottom of the upper figures show the Rabi potential density $V_R$, and the magenta (cyan) makers indicate
the position of $u$ vortex ($d$ vortex) from $\tau =0$ to $200$. 
$u =  2u_{12} = 1$ and $\eta = 0.01$. The figures at the second row are the upper figures seen from the top. 
}
\label{fig:v10v01compare}
\end{center}
\end{figure}

\begin{figure}[h]
\begin{center}
\includegraphics[width=8cm]{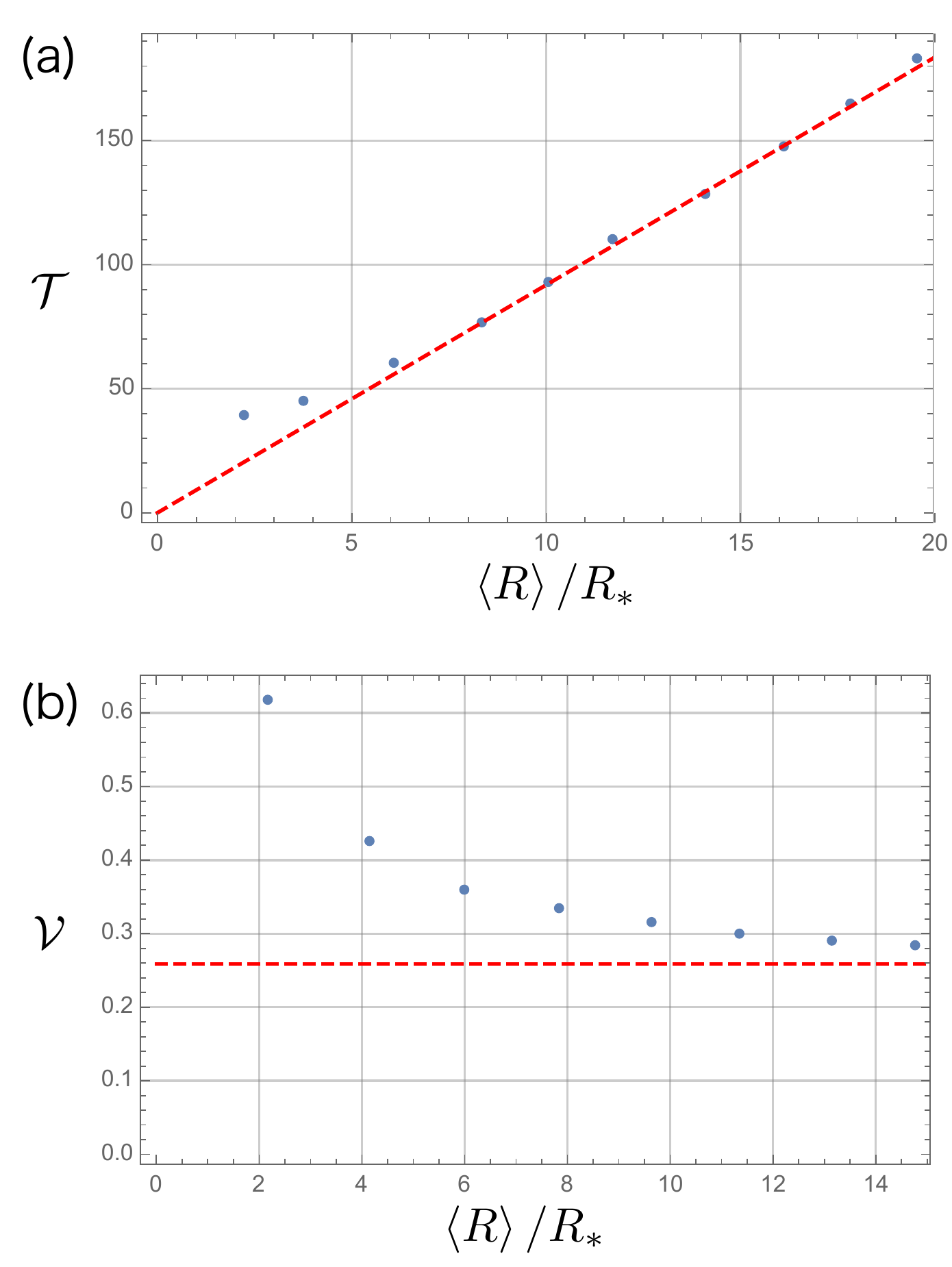}
\caption{
(a) The period ${\cal T}$ of precession of $ud$ as a function of the mean size $\left<R\right>$. 
The dashed line is the asymptotic linear function
with $\alpha = 0.85$.
(b) The velocity ${\cal V}$ of $\bar uu$ as a function of the mean size $\left<R\right>$. The
red dashed line corresponds to the asymptotic constant value 
with $\alpha = 0.91$. 
}
\label{fig:eta_size_period_velocity}
\end{center}
\end{figure}

Once the size of baryon deviates from the equilibrium $R_0$, it starts to rotate,
since the interaction is repulsive (attractive) for $R < R_0$ ($R > R_0$).
Let us show several examples with three different sizes i) $R < R_0$, ii) $R=R_0$ and iii) $R>R_0$.
Fig.~\ref{fig:v10v01compare} shows typical time evolutions of the baryonic molecules. 
As can be clearly seen, the size of molecule affects the rotation. 
The baryon of $R < R_0$ ($R > R_0$) rotates counterclockwise 
(clockwise).
When the molecule size is sufficiently large as $R > R_*$ (but less than the critical length $R_c$),  
the soliton tension dominates the inter-vortex force.
The rotation period ${\cal T}$ of a baryonic molecule is 
asymptotically given by 
${\cal T}  \to  \frac{\pi^2}{4\alpha \sqrt{\eta}} \rho$.
We numerically confirm that ${\cal T}$ indeed approaches the asymptotic behavior as shown in Fig.~\ref{fig:eta_size_period_velocity}(a). 

On the other hand, a mesonic molecule moves at a constant velocity.
The velocity of a mesonic molecule
is asymptotically given by 
${\cal V} \to \frac{8\alpha\sqrt{\eta}}{\pi}$.
Indeed,
${\cal V}$ asymptotically approaches to the asymptotic value as $R$ being increased, as can be seen in Fig.~\ref{fig:eta_size_period_velocity}(b).

\begin{figure}[h]
\begin{center}
\includegraphics[width=8cm]{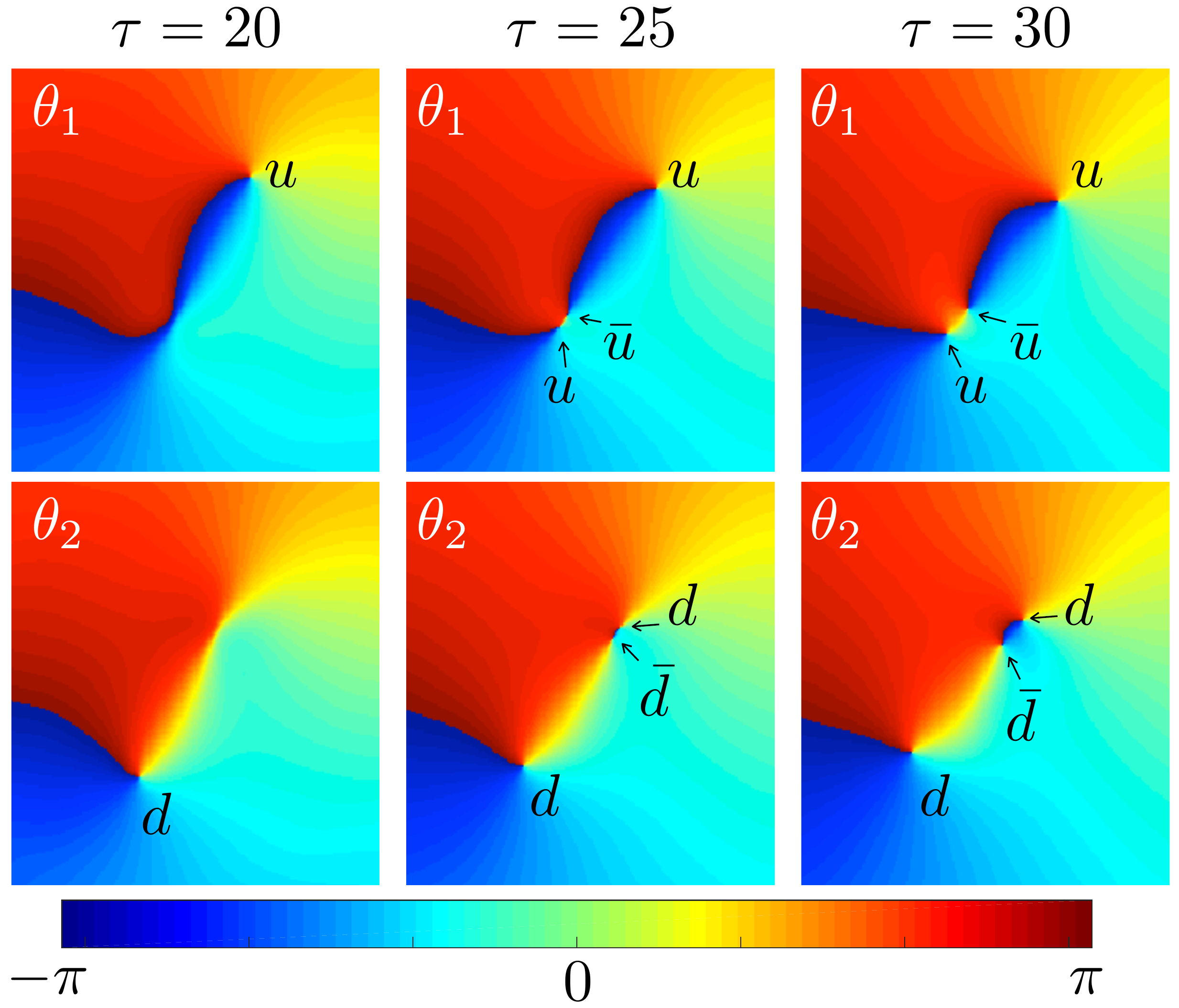}
\caption{ 
The phases $\theta_{1,2}$ of $\Psi_{1,2}$ for the motion of Fig.~3. Three snapshots at $\tau = 20,25,30$, at which 
the string fragmentation occurs, are shown. The plot regions are $\vec y \in [-20,20]^2$.
}
\label{fig:phase_baryon}
\end{center}
\end{figure}

When we initially set the size of molecule longer than the critical size $R_c$, the molecule 
breaks up with pair creation of vortex and anti-vortex.
The phases of $\Psi_1$ and $\Psi_2$ clearly show 
a pair creation phenomenon, as in Fig.~\ref{fig:phase_baryon}, 
in which fragmentation of a long baryon $ud$ ($R_{\rm ini} = 24R_*$) is shown. 
From this figure, we observe that two pairs, $u\bar u$ and $d\bar d$, are dynamically created and the soliton is chopped. 
As a consequence, the initial long $ud$ baryon splits up into three short molecules:
two short $ud$ baryons and a semi-long anti-baryon $\bar d\bar u$. Thus, the baryon number is preserved under this process. 

\begin{figure*}[h]
\begin{center}
\includegraphics[width=15cm]{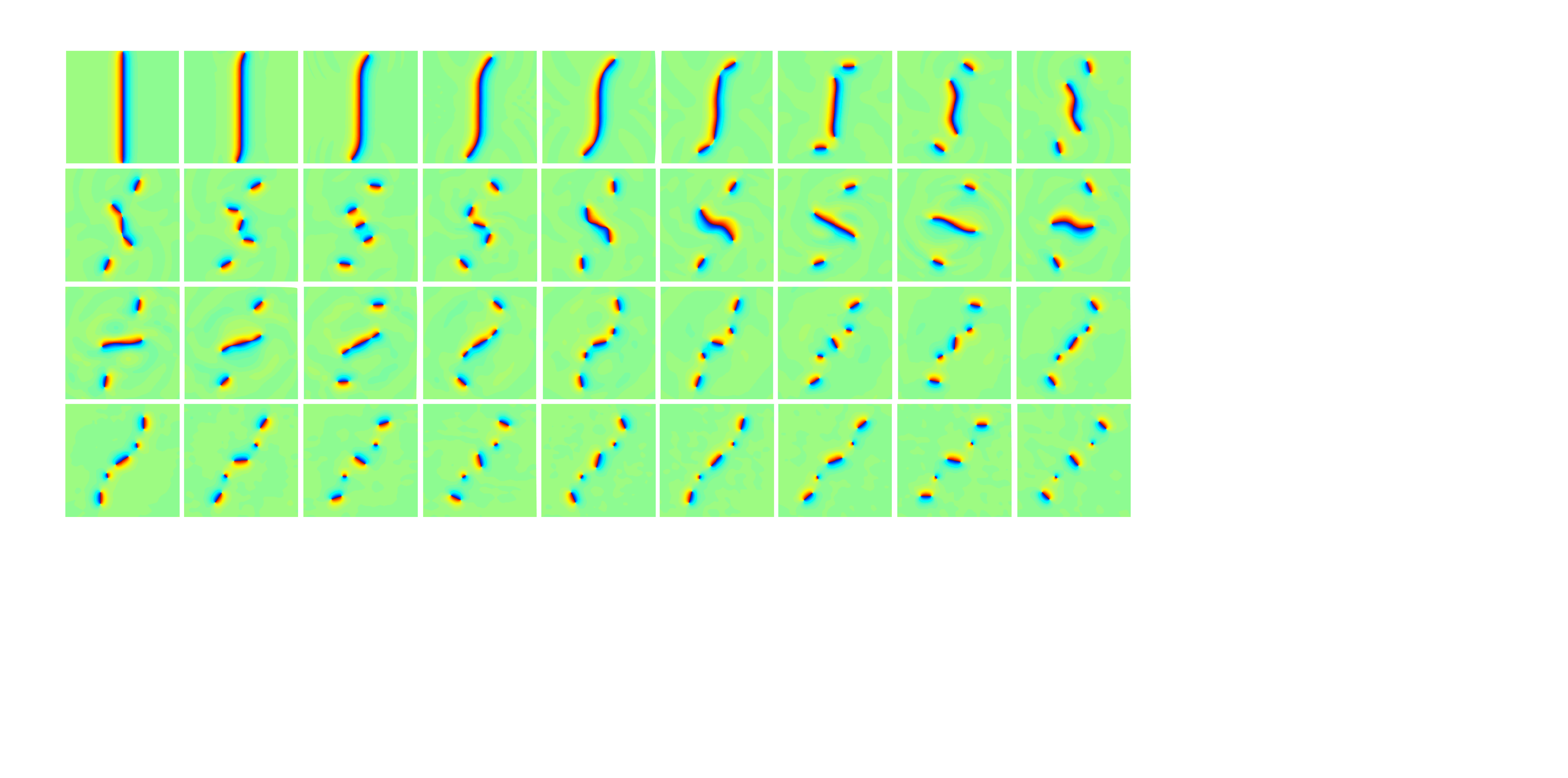}\\\ \\
\includegraphics[width=15cm]{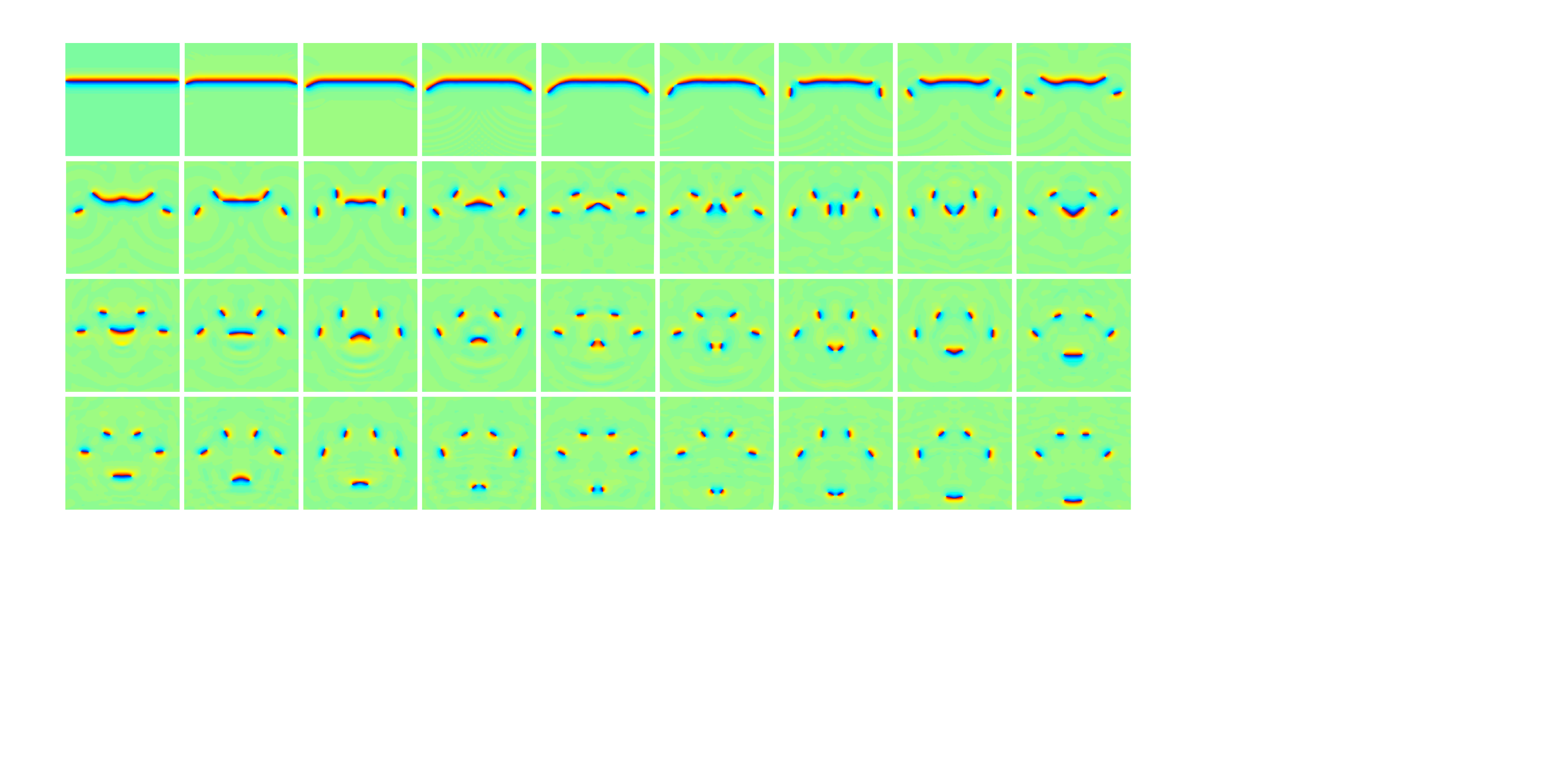}
\caption{
(The upper panels) A long baryon with $R_{\rm ini} = 44 R_*$ breaks up into 3 baryons and 2 anti-baryons. 
(The lower panels) A long meson with $R_{\rm ini} = 60 R_*$ breaks up into 2 baryons, 2 anti-baryons and 1 meson.
Real time evolutions for $\tau = 0$\,-\,$175$ with a step $\Delta \tau = 5$.
}
\label{fig:baryon_5}
\end{center}
\end{figure*}

When we put a longer baryon ($R_{\rm ini} = 44 R_*$) at initial time, it splits up into five small pieces as shown in Fig.~\ref{fig:baryon_5}(top).
Fragmentation occurs in multifolds.
Initially, the very long $ud$ baryon releases two short $ud$ baryons from its edges remaining a long anti-baryon ($\bar d \bar u$) at center as
$ud \to ud+\bar d\bar u+ud$.
Soon after the first fragmentation, the two short anti-baryons are separated from the long anti-baryon, leaving another $ud$ baryon at the center
as $ud+\bar d\bar u+ud+\bar d\bar u+ud$. 
Then the three baryons rotate clockwise and the two anti-baryons rotate counterclockwise. After a while the configuration becomes
$ud+\bar u\bar d+du+\bar u\bar d+ud$, and then the three molecules inside
coalesce back into a long anti-baryon as
$ud+\bar u\bar d+ud$.  Finally, the long anti-baryon $\bar u\bar d$ at the center again breaks up into the five pieces as
$ud+\bar u\bar d+du+\bar u\bar d+ud$.
The baryon number is preserved at any stage. 

Similarly, a very long meson ($R_{\rm ini} = 60 R_*$) at initial time breaks up into five or more small molecules as shown in the lower panels of 
Fig.~\ref{fig:baryon_5}.
As for baryons, the fragmentation gradually occurs from the ends of the molecule. Initially, the molecule is $\bar u u$ and it splits into 
$\bar u \bar d$+$d\bar d$+$d u$. At the second stage, it further breaks up into $\bar u \bar d$+$du$+$\bar u u$+$\bar u\bar d$+$d u$.
Then, it breaks up into six pieces as $\bar u \bar d$+$du$+$\bar u\bar d$+$du$+$\bar u\bar d$+$d u$.
The meson at the center moves straight downward 
with showing the flavor oscillations
as
$\bar u u$ $\leftrightarrow$ $\bar u\bar d$+$du$
$\leftrightarrow$ $\bar d\bar u$+$ud$ $\leftrightarrow$ $\bar d d$.

\section{Baryons and mesons in QCD and BEC}
\subsection{Baryons and mesons in $SU(2)$ QCD}
Here, we consider $SU(2)$ QCD with two quarks (flavors) 
$u$- and $d$-quarks, and compare states in this theory 
with those of BECs.
In the massless limit, these quarks belong to a doublet of the $SU(2)$ 
flavor symmetry.
Each quark also belongs to a color doublet as $u = (u_r,u_g)$ and $d=(d_r,d_g)$,
where $r$ and $g$ are color indices. 
The color singlet states composed of $u$- and $d$-quarks
are a triplet baryon and a singlet baryon of the $SU(2)$ flavor symmetry 
as described below.
In detail, the triplet baryon is symmetric in the flavor index 
and antisymmetric in the color index: 
\beq
\left(
\begin{array}{c}
u_ru_g-u_gu_r \cr
u_rd_g + d_ru_g - ( u_gd_r+d_gu_r) \cr
d_rd_g-d_gd_r
\end{array}\right), 
   \label{eq:baryon3}
\eeq 
while the singlet baryon is antisymmetric both for the flavor and color indices:
\beq
u_rd_g -d_r u_g - (u_g d_r - d_g u_r). \label{eq:baryon1}
\eeq
Taking into account the fact that spin of quark is $1/2$, the triplet should be spin 1, and the singlet should be spin 0.

Next, let us construct mesons.
There are triplet meson, given by 
\beq
\left(
\begin{array}{c}
  u_r \bar d_r - u_g \bar d_g \cr 
  u_r\bar u_r - d_r\bar d_r - (u_g\bar u_g - d_g\bar d_g) \cr 
  d_r\bar u_r - d_g\bar u_g 
\end{array}\right),
\eeq 
and  a singlet meson, given by
\beq 
 u_r \bar u_r + d_r \bar d_r - (u_g \bar u_g + d_g\bar d_g).
\eeq
Here, we use the notation $\bar u_r = \overline{u_g}$, etc.

\subsection{Baryons and mesons in bosonic $SU(2)$ QCD}

Let us consider $SU(2)$ QCD with {\it bosonic} quarks,
since the charged particles concerned in this work are 
bosons (spin 0) which are dual to 
the HQVs in two component BECs. 
If $u$ and $d$ are bosons, the components of triplet baryon corresponding to 
Eq.~(\ref{eq:baryon3}) vanish, while 
the singlet baryon corresponding to Eq.~(\ref{eq:baryon1}) 
survives as $u_rd_g -d_r u_g$.
Unlike the baryons, both the triplet and singlet mesons survive
even if $u$ and $d$ are bosons.
Thus, what we observe in spectra of the bosonic composite states are
the singlet baryon, triplet and singlet mesons. The number of the composites is $1+3+1 = 5$.

\subsection{Baryons and mesons in coherently coupled BECs}
Let us compare these with those in coherently coupled BECs
which are the $ud$ baryon and $u \bar u$ and $d \bar d$ 
as found in this work. 
In order to obtain the states in BECs, 
let us truncate the off-diagonal components $u_g$ and $d_r$ 
in the bosonic QCD.
The two component BECs realize only $U(1)$ subgroups of the $SU(2)$ color and $SU(2)$ flavor symmetries.
Then, 
the singlet baryon is 
\beq 
 u_rd_g,
\eeq
while 
the 
triplet and singlet mesons reduce to 
\beq
\left(
\begin{array}{c}
  0 \cr u_r\bar u_r + d_g\bar d_g \cr 0 
\end{array}\right),
\quad
 u_r \bar u_r - d_g\bar d_g,
\eeq  
 respectively.
They are nothing but what we observed in 
coherently coupled two-component BECs.

\clearpage

\newcommand{\J}[4]{{\sl #1} {\bf #2} (#3) #4}
\newcommand{\andJ}[3]{{\bf #1} (#2) #3}
\newcommand{\AP}{Ann.\ Phys.\ (N.Y.)}
\newcommand{\MPL}{Mod.\ Phys.\ Lett.}
\newcommand{\NP}{Nucl.\ Phys.}
\newcommand{\PL}{Phys.\ Lett.}
\newcommand{\PR}{ Phys.\ Rev.}
\newcommand{\PRL}{Phys.\ Rev.\ Lett.}
\newcommand{\PTP}{Prog.\ Theor.\ Phys.}
\newcommand{\hep}[1]{{\tt hep-th/{#1}}}


\begin{thebibliography}{100}
\bibitem{footnote2}
$u$ and $d$ quarks in QCD
have a color label as $u_a$, $d_a$ ($a = r,g$).
We identify $u_r$ ($d_g$) with the $\theta_1$ ($\theta_2$) vortex 
while $u_g$ and $d_r$ are truncated by the diagonal approximation.
The $U(1)$ subgroup of $SU(2)$ color group remains 
manifest, which we identify $U(1)_R$. 
See Appendix B for spectra of mesons and baryons
constructed by the $u_r$ and $d_g$ only. 
  
\if0
\end{thebibliography}
\end{document}